\documentclass[iop,revtex4,onecolumn]{emulateapj}

\newcommand{\nolinkurl}{\url}

\begin{document}

\title{On the most reliable value of the Galactic aberration constant}
\author{Zinovy Malkin}
\affiliation{Pulkovo Observatory, St.~Petersburg 196140, Russia}
\email{zmalkin@zmalkin.com}

\begin{abstract}
Galactic aberration (GA) is a small effect in proper motions of celestial objects
with an amplitude of about 5 uas/yr already noticeable in highly accurate astrometric
observations such as VLBI and $Gaia$.
However accurate accounting for this effect faces difficulty caused by the uncertainty
in the GA amplitude (GA constant).
Its estimates derived from VLBI and $Gaia$ data processing differ significantly,
so it would be very desirable to involve another independent method to solve the problem
of inconsistency between these two methods.
Such a method, that we consider in this paper, is using determination of the Galactic rotation
parameters by methods of stellar astronomy.
The result obtained in this study showed that the GA constant estimate obtained
from stellar astronomy is closer to the estimate obtained from $Gaia$.
\end{abstract}

%\keywords

\maketitle

%%%%%%%%%%%%%%%%%%%%%%%%%%%%%%%%%%%%%%%%%%%%%%%%%%%%%%%%%%%%%%%%%%%%%%%%%%%%%%%%%%%%%%%%%%%%%%

\section{Introduction}

The precision of the most accurate modern catalogs of extragalactic source positions, such as ICRF
(International Celestial Reference Frame) in radio and $Gaia$ in optics, is at the level of 10 $\mu$as.
At such a high level, even tiny effects can affect the source positions derived from radio and optical
observations.
One of these effects is Galactic aberration in proper motions (GA) caused by the centripetal acceleration
of the solar system (more accurately, local standard of rest (LSR)), toward the Galactic center
\citep{Kovalevsky2003,Malkin2014a}.
The impact of the GA on the apparent source motion depends on the source coordinates and has maximum 
amplitude of $\approx \! 5$~$mu$as/yr, which can be called GA constant, hereafter denoted as $A$,
see Section~\ref{sect:modeling} for details.
A discussion on the best estimate of the GA constant is given in \citet{Malkin2014a,MacMillan2019}
The authors of both papers showed a significant disagreement between its value derived from VLBI absolute
astrometry ($A =5.8 \pm 0.3$~$\mu$as/yr) and from the Galactic rotation parameters obtained by methods
of stellar astronomy ($A = (4.9 \ldots 5.0) \pm (0.2 \ldots 0.3)$~$\mu$as/yr).
The $Gaia$ observations gave $A = 5.05 \pm 0.35$~$\mu$as/yr \citep{Klioner2021},
which is closer to the value obtained from stellar astronomy.
In this paper, the latter estimate was revised using recent results of simultaneous determination of the Galactic
rotation parameters, such as Galactocentric distance of the Sun and Galaxy rotational velocity.

%%%%%%%%%%%%%%%%%%%%%%%%%%%%%%%%%%%%%%%%%%%%%%%%%%%%%%%%%%%%%%%%%%%%%%%%%%%%%%%%%%

\section{GA modeling}
\label{sect:modeling}

The apparent proper motion of extragalactic sources caused by the GA is given by \citep{Kovalevsky2003,Malkin2014a}:
\begin{equation}
\begin{array}{rcl}
\mu_l \cos b &=& -A \sin l \,, \\
\mu_b &=& -A \cos l \sin b \,, \\
\end{array}
\label{eq:galactic}
\end{equation}
where $l$ and $b$ are Galactic longitude and latitude of the object, respectively, and $A$ is the GA constant,
which depends on the Galactic rotation parameters as \citep{Malkin2014a}
\begin{equation}
A = \frac{V_0 \Omega_0}{c} = \frac{R_0 \Omega_0^2}{c} = \frac{V_0^2}{R_0 \, c} \,,
\label{eq:Aconst}
\end{equation}
where $R_0$ is the Galactocentric distance of the Sun, $V_0$ and $\Omega_0$ are the linear and angular velocities,
respectively, of the LSR's revolution around the Galactic center, and $c$ is the speed of light.

%%%%%%%%%%%%%%%%%%%%%%%%%%%%%%%%%%%%%%%%%%%%%%%%%%%%%%%%%%%%%%%%%%%%%%%%%%%%%%%%%%%%%%%%%%%%%%

\section{GA constant from stellar astronomy}
\label{sect:comparison_result}

To obtain the most reliable estimate of the GA constant $A$ all the recent estimates of the Galactic
rotation parameters were used.
Two strategies can be used for this purpose.
One can evaluate $R_0$, $V_0$, and $\Omega_0$ independently from the papers where these values were determined
together or separately.
Such an approach was used in previous papers of the author, such as \citet{Malkin2011,Malkin2014a,Malkin2014b},
and yielded $A$ values from $4.9 \pm 0.4$ to $5.0 \pm 0.3$~$\mu$as/yr.
However, in many cases, the estimates of the Galactic rotation parameters are not independent in the original
papers used to get these $A$ estimates.  
To make this estimate more systematically independent, only papers in which $R_0$ and LSR velocity have been
estimated simultaneously were used here.
The result of this analysis is presented in Table~\ref{tab:a_stellar}. 
The  mean value of all estimates listed in the last column of Table~\ref{tab:a_stellar} was computed using
extended error handling considering both the random errors and scatter of the input data \citep{Malkin2013}.
It was found to be $A = 4.7 \pm 0.1$~$\mu$as/yr.

\begin{table}[ht]
\centering
\caption{Results of computation of the GA constant $A$ from stellar astronomy.}
\label{tab:a_stellar}
\begin{tabular}{lcccc}
\hline
Reference                & $R_0$ [kpc] & $V_0$ [km/s] & $\Omega_0$ [km/s/kpc] & $A$ [$\mu$as/yr] \\
\hline                                                                              
\citet{McMillan2017} (a) & 8.20 $\pm$ 0.09   & 232.8 $\pm$ 3.0    &                             & 4.65 $\pm$ 0.13 \\
\citet{McMillan2017} (b) & 7.97 $\pm$ 0.15   & 226.8 $\pm$ 4.2    &                             & 4.54 $\pm$ 0.19 \\
\citet{Rastorguev2017}   & 8.24 $\pm$ 0.12   & 236.5 $\pm$ 7.0    &                             & 4.78 $\pm$ 0.29 \\
\citet{Xu2018}           & 8.35 $\pm$ 0.18   & 223.0 $\pm$ 13~    &                             & 4.19 $\pm$ 0.50 \\
\citet{Reid2019}         & 8.15 $\pm$ 0.15   & 236.0 $\pm$ 7.0    &                             & 4.81 $\pm$ 0.30 \\
\citet{VERA2020}         & 7.92 $\pm$ 0.16   &                    & 28.63 $\pm$ 0.26            & 4.57 $\pm$ 0.12 \\
\citet{Bobylev2020}      & 8.15 $\pm$ 0.12   & 236.4 $\pm$ 4.4    &                             & 4.82 $\pm$ 0.19 \\
\citet{Bobylev2021}      & 8.27 $\pm$ 0.10   & 240.0 $\pm$ 3.0    &                             & 4.90 $\pm$ 0.14 \\
\end{tabular}
\end{table}

%%%%%%%%%%%%%%%%%%%%%%%%%%%%%%%%%%%%%%%%%%%%%%%%%%%%%%%%%%%%%%%%%%%%%%%%%%%%%%%%%%%%%%%%%%%%%%%%%%%

\section{Conclusion}
\label{sect:conclusion}

In this paper, a new estimate of the GA constant, $A = 4.7 \pm 0.1$~$\mu$as/yr, was obtained.
This estimate is much closer to the estimate obtained by the $Gaia$ team \citep{Klioner2021},
$A =5.05 \pm 0.35$~$\mu$as/yr, than to the estimated based on the geodetic VLBI,
$A =5.8 \pm 0.3$~$\mu$as/yr.
This allows us to assume that the geodetic VLBI-based $A$ estimate may give an overestimated $A$ value,
while the estimate obtained here, as well in previous papers \citet{Malkin2011,Malkin2014a,Malkin2014b},
agrees well with the $Gaia$ estimate within the formal errors.

It should be noted that the uncertainty of the $A$ estimate obtained in this paper may be underestimated
because many of the cited papers are based on the same observations and, therefore, are statistically dependent.

%%%%%%%%%%%%%%%%%%%%%%%%%%%%%%%%%%%%%%%%%%%%%%%%%%%%%%%%%%%%%%%%%%%%%%%%%%%%%%%%%%%%%%%%%%%%%%%%%%%

\bibliography{ga_a}
\bibliographystyle{aasjournal}

\end{document}